\documentclass[conference]{IEEEtran}

\usepackage{cite}
\usepackage{amsmath,amssymb,amsfonts}
\usepackage{algorithmic}
\usepackage{graphicx}
\usepackage{textcomp}
\usepackage{xcolor}
\def\BibTeX{{\rm B\kern-.05em{\sc i\kern-.025em b}\kern-.08em
    T\kern-.1667em\lower.7ex\hbox{E}\kern-.125emX}}

\usepackage{tikz}

\begin{document}

\title{Metadata-Private Voice Communication over Tor Using Post-Quantum Key Encapsulation\\}

\author{\IEEEauthorblockN{1\textsuperscript{st} Noah Zeinali Yazdi}
\IEEEauthorblockA{\textit{Independent Researcher} \\
Wiesbaden, Germany \\
ORCID: 0009-0005-1723-2339}
\and
\IEEEauthorblockN{2\textsuperscript{nd} Samuel Zeinali Yazdi}
\IEEEauthorblockA{\textit{Independent Researcher} \\
Wiesbaden, Germany \\
ORCID: 0009-0007-8518-0635}
}

\maketitle

\begin{abstract}
Encrypted voice communication systems commonly protect call content but expose significant metadata, including the network identities of both parties, call timing, and traffic patterns. 
Simultaneously, advances in quantum computing threaten the asymmetric primitives underlying conventional key exchange, creating a harvest-now, decrypt-later risk for long-lived sensitive communications. Existing systems address at most one of these concerns: secure VoIP applications such as Signal and Wire rely on centralized infrastructure and classical cryptography, while anonymous communication tools built on the Tor network support only text messaging and lack post-quantum key encapsulation.

This paper presents SAPEC (Secure Anonymous Peer Exchange Communicator), a peer-to-peer 
voice communication system that simultaneously addresses anonymity, decentralization, and 
quantum resistance. SAPEC routes all traffic exclusively through Tor hidden services, 
concealing the IP addresses of both parties, and augments the standard TLS handshake with 
a custom protocol layer that performs ML-KEM-1024 key establishment, a settle on AES-256 symmetrical keys and mutual Dilithium identity verification before any audio is transmitted. All audio 
is subsequently encrypted under AES-256. A removable Crypto Ignition Key prevents identity key material from being stored on the device.

Empirical evaluation across 45 independently circuit-rebuilt test runs yielded a median 
one-way audio delay of 355 milliseconds and a median connection setup time of 7.61 seconds on same-region Tor circuits, demonstrating that interactive voice communication over onion routing is practically achievable. The system has been deployed on three hardware platforms: a desktop client, a Raspberry Pi 5 embedded in a retrofitted desk telephone, and a Raspberry Pi Zero 2W pager. To the best of our knowledge, SAPEC is the first working prototype to combine post-quantum key encapsulation, anonymous overlay routing, and real-time voice communication on consumer embedded hardware.
\end{abstract}
\begin{IEEEkeywords}
secure telephony, quantum security, onion-routing, tor, metadata minimization, anonymous communication
\end{IEEEkeywords}

\section{Introduction}
Achieving private and anonymous communication is an ongoing challenge. Specifically regarding voice communication, despite end-to-end encryption that is commonly used, there is significant exposure of metadata, i.e., further information attached to the encrypted data\cite{async}. Furthermore, besides metadata, quantum computing puts asymmetric encryption itself at risk: With quantum-algorithms, conventional RSA and ECC public-key cryptography could be attacked through Shor's algorithm, significantly reducing its computational hardness\cite{breakcrypt}. While quantum-computing at this scale is not yet practicably available, confidential data should be resistant to future threats by design\cite{nistpq}. Moreover, centralized infrastructure puts providers of services at risk of being targeted by cyberattacks\cite{target}.

Existing communication systems can be evaluated with respect to anonymity, decentralization, and resistance against quantum attacks (see Section~\ref{sec:relatedwork}). 

Our Research with the development name \textit{SAPEC} - Secure Anonymous Peer Exchange Communicator is, to the best of our knowledge, the first system to combine post-quantum key encapsulation, anonymous overlay routing, and metadata cascading into a working voice communication prototype on consumer embedded hardware.

The remainder of this paper is structured as follows. 
Section~\ref{sec:background} provides the technical background on ML-KEM, Tor onion routing, 
and TLS necessary to understand the SAPEC design.

Section~\ref{sec:relatedwork} positions SAPEC 
within the landscape of existing secure VoIP systems, anonymous communication tools, and 
post-quantum messaging research, and identifies the gap that SAPEC occupies.

Section~\ref{sec:design} describes the system design in full, covering the threat model, 
overall architecture, the modified TLS handshake that constitutes the primary protocol 
contribution, and the complementary metadata minimization mechanisms.

Section~\ref{sec:implementation} details the hardware and software implementation across the 
CipherLink backend, the SAPEC Desk Phone, and the SAPEC Pager.

Section~\ref{sec:evaluation} presents measurements of latency, RTT and
connection setup time.

Section~\ref{sec:discussion} interprets these results, acknowledges current limitations, and outlines directions for future work. 
Section~\ref{sec:conclusion} concludes the paper.

\section{Background}
\label{sec:background}

\subsection{ML-KEM / Kyber}

ML-KEM (Module Lattice-based Key Encapsulation Mechanism), standardized by NIST as FIPS 203 \cite{fips203} and previously known under the name Kyber, is a post-quantum key encapsulation mechanism whose security is grounded in the hardness of the Module Learning With Errors (MLWE) problem. At its core, the algorithm operates over a high-dimensional lattice: a structured grid of points generated by randomly sampled polynomial functions. The cryptographic challenge reduces to the Closest Vector Problem (CVP), given a point placed near but not on a lattice point, an adversary must determine which lattice point is nearest. Unlike factorization-based schemes such as RSA, no efficient quantum algorithm is known to solve this problem, rendering ML-KEM resistant to attacks by Shor's algorithm.

The practical security of ML-KEM scales with the dimensionality of the underlying lattice. The standard defines three parameter sets: ML-KEM-512, ML-KEM-768, and ML-KEM-1024, corresponding to 512, 768, and 1024 dimensions respectively. An adversary attempting a brute-force search must enumerate candidate lattice points across all dimensions simultaneously, a task that remains computationally infeasible even for large-scale quantum hardware. Coordinates are not restricted to integers but extend to values modulo a prime $q = 3329$, further enlarging the search space. This combination of high dimensionality, non-integer coordinates, and an undisclosed lattice basis constitutes the hardness guarantee that makes ML-KEM suitable for long-term cryptographic security \cite{fips203,kyber}.

\subsection{Tor Onion Routing}

The Tor network is a volunteer-operated overlay network designed to provide anonymous communication over the public Internet \cite{toroverview}. Each participant is identified not by an IP address but by a 56-character onion address\cite{onionaddress} derived from a public key, functioning analogously to a conventional domain name. When a client establishes a connection through Tor, traffic is routed through a sequence of typically three relays: a guard node, one or more intermediate relays, and an exit node. Each relay in the chain maintains knowledge only of its immediate predecessor and successor, so no single node can correlate both the origin and destination of a connection. Encryption is applied in layers (hence the onion metaphor) with each relay decrypting only the outermost layer to determine the next hop\cite{toroverview}.

For hidden services, no exit node is required; both parties rendezvous inside the Tor network itself, ensuring that neither endpoint's IP address is exposed to the other or to any relay. Standard circuit depth is three to five relays; circuits spanning relays across multiple geographic regions can introduce latencies of several seconds, though typical same-continent paths yield latencies compatible with real-time audio \cite{torlatency}. This architecture makes Tor the appropriate transport layer for a voice communication system whose threat model includes IP-based deanonymization attacks against either party.

\subsection{Transport Layer Security}

Transport Layer Security (TLS) is a widely used protocol for establishing authenticated, confidential channels over the Internet \cite{tlsrfc}. A TLS handshake begins with version and cipher suite negotiation, followed by an asymmetric key exchange, typically Elliptic Curve Diffie-Hellman Ephemeral (ECDHE), that establishes a shared session secret from which symmetric session keys are derived. Server authentication is provided by a certificate chain rooted in a trusted Certificate Authority. Once the handshake completes, all application data is encrypted under a symmetric cipher such as AES-256-GCM. TLS thus separates the concerns of identity verification, key agreement, and bulk data confidentiality into a well-defined, widely implemented protocol stack. However, the asymmetric primitives employed in the standard handshake are not quantum-resistant\cite{tlsrfc, tlsmicrosoft}, motivating the protocol extension described in Section~\ref{sec:design}.

\section{Related Work}
\label{sec:relatedwork}

Existing secure communication systems can be evaluated along three axes relevant to the SAPEC threat model: anonymity of both communicating parties, decentralization of infrastructure, and resistance to attacks by quantum-capable adversaries. No prior system simultaneously satisfies all three requirements for real-time voice communication on consumer embedded hardware per our knowledge.

\subsection{Secure VoIP Systems}

Signal and Wire represent the current state of the art in encrypted voice communication for consumer devices\cite{signalwire}. Both implement the Double Ratchet Algorithm over an authenticated key exchange, providing forward secrecy and break-in recovery for their respective messaging layers, and both extend this protection to voice calls via SRTP encrypted with keys derived from the same ratchet mechanism \cite{keysignal, keysignal2, keysignal3, keywire}.From a content confidentiality perspective, these systems have been available for the longest period among the services evaluated and may therefore, in our assessment, be regarded as comparatively more established in the market.

However, both systems fail to satisfy the anonymity and decentralization requirements of the SAPEC threat model. Signal and Wire route all traffic including call signaling and media through centralized infrastructure operated by their respective organizations\cite{keywire, signallegal}, meaning that metadata such as call timing, duration, and the network addresses of both participants may be visible to the service provider. Registration requires a phone number, binding identity to a telecommunications carrier record\cite{signallegal, keywire}. Furthermore, neither system employs post-quantum key encapsulation in its handshake; their asymmetric primitives remain vulnerable to a harvest-now, decrypt-later attack by a quantum-capable \cite{HNDL}. SAPEC addresses all three shortcomings by routing calls exclusively over Tor hidden services, requiring no persistent identifier beyond an onion address, and replacing the classical key exchange with ML-KEM-1024.

\subsection{Anonymous Communication Systems}

Briar and Cwtch, represent a family of privacy-focused communication tools that route traffic over the Tor network and dispense with centralized servers entirely. Cwtch operates purely over Tor hidden services and requires no account registration, providing strong anonymity and decentralization \cite{cwtchsecurity}. Cwtch extends this model with multi-party server-optional group messaging, and Briar additionally supports Bluetooth and WiFi-Direct transports for use in infrastructure-denied environments \cite{briarsecurity, cwtchsecurity}.

Despite their strong anonymity properties, these systems are designed exclusively for text messaging and do not support real-time audio transmission. None of the above projects attempt low-latency audio delivery over onion-routed connections. Moreover, none incorporate post-quantum key encapsulation in their handshakes, leaving long-term identity keys exposed to future quantum attacks\cite{briarsecurity, cwtchsecurity}. SAPEC demonstrates that real-time voice communication over Tor is feasible at bitrates between 10\,kbps and 64\,kbps with latencies remaining within acceptable bounds for conversational use, and augments the anonymity guarantees of onion routing with an ML-KEM layer over the TLS handshake.

\subsection{Post-Quantum Secure Messaging Research}

Several research prototypes have explored the integration of post-quantum primitives into secure 
messaging protocols. The PQXDH specification, adopted by Signal as a replacement for X3DH, 
incorporates CRYSTALS-Kyber alongside X25519 to provide a hybrid key agreement that is secure if either primitive remains unbroken \cite{keysignal4}. Academic work has similarly investigated 
post-quantum adaptations of the Double Ratchet and group key agreement schemes \cite{pqratchet}.

These contributions, however, operate at the application-layer messaging level and inherit the 
centralized infrastructure and metadata exposure of their parent systems. They also exclusively 
target text and asynchronous messaging rather than real-time voice\cite{keysignal4, pqratchet}. To the best of our knowledge, no prior published system combines post-quantum key encapsulation with anonymous overlay routing for the specific purpose of real-time voice communication on consumer embedded hardware. SAPEC occupies this gap, providing a working prototype that achieves all three properties: anonymity, decentralization, and quantum resistance, within the latency constraints imposed by interactive telephony.

\section{System Design}
\label{sec:design}

SAPEC is a peer-to-peer voice communication system designed to provide simultaneously anonymous, decentralized, and post-quantum-secure telephony on consumer embedded hardware. Before describing its constituent parts, it is useful to characterize the system as a whole: two parties, each identified solely by a Tor hidden-service address, establish a bidirectional audio channel whose setup is protected by a custom protocol layer above TLS, whose session keys are derived with the assistance of ML-KEM-1024, and whose traffic is routed exclusively through the Tor network so that neither party's IP address is revealed to the other or to any intermediary (after the guard node) \cite{toroverview}. The software backend, CipherLink, is hardware-agnostic and has been deployed on a conventional desktop, a Raspberry Pi 5 embedded in a retrofitted desk telephone, and a Raspberry Pi Zero 2W in a portable pager form factor.

\subsection{Threat Model}

The adversary considered by SAPEC is a passive or active network observer with significant 
resources. At the network layer, the adversary is assumed to be capable of observing all traffic on the public Internet, recording encrypted ciphertext for offline decryption once quantum hardware becomes available to excecute a so-called harvest-now, decrypt-later attack. The adversary can observe packet timing, packet sizes, and connection patterns between IP addresses, and may attempt to correlate these observations to deanonymize either communicating party. At the protocol layer, the adversary may attempt replay attacks, injecting or re-transmitting captured audio packets in an attempt to disrupt a session or trigger erroneous decryption. Physical access to a seized device is also considered: if identity keys are stored unprotected on disk, a device seizure reveals the user's persistent onion address and contact history.

SAPEC does not consider an adversary who has compromised the Tor network globally, nor one who has obtained administrative access to both endpoints simultaneously. The system is accordingly not designed to resist a global passive adversary performing end-to-end traffic correlation across all Tor relays. Within the stated threat model, SAPEC aims to protect: (1) the IP addresses and network identities of both parties; (2) the content of voice calls against current and future quantum adversaries; (3) session metadata, including call timing and packet-level traffic patterns; and (4) long-term identity keys against physical device seizure.

\subsection{System Overview}

\begin{figure}[htbp]
\centerline{\includegraphics[width=\columnwidth]{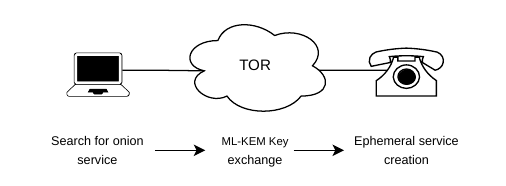}}
\caption{SAPEC system architecture. Both endpoints run CipherLink behind Tor hidden services. 
No IP addresses are exchanged; all signaling and audio traverse the Tor network. The initial connection step is 
established via ML-KEM public key pairs stored on a removable CIK device.}
\label{fig:architecture}
\end{figure}

The top-level architecture of SAPEC is shown in Fig.~\ref{fig:architecture}. Each endpoint runs an instance of CipherLink, a Python-based backend \cite{python} that manages the Tor hidden service, the modified TLS handshake, audio capture and playback, and the nonce counter state required for replay detection. The Tor daemon runs locally on each device and exposes a SOCKS5\cite{socks5} proxy through 
which CipherLink directs all outbound connections, while simultaneously advertising a hidden 
service on a device-specific onion address derived from the user's identity key pair.

A user is reachable at their onion address as soon as the hidden service descriptor has propagated to the Tor distributed hash table, a process that typically completes within 30 to 60 seconds of daemon startup No central server, directory service, or registration authority is involved at any point\cite{toroverview}. Contacts are stored locally as a JSON file phonebook mapping human-readable names to onion addresses. Identity key material is never stored on the device's internal storage; instead, it resides on a removable USB device designated the Crypto Ignition Key (CIK), described further in Section~\ref{subsec:metadata}.

\subsection{Modified TLS Handshake}

\begin{figure}[htbp]
\centerline{\includegraphics[width=\columnwidth]{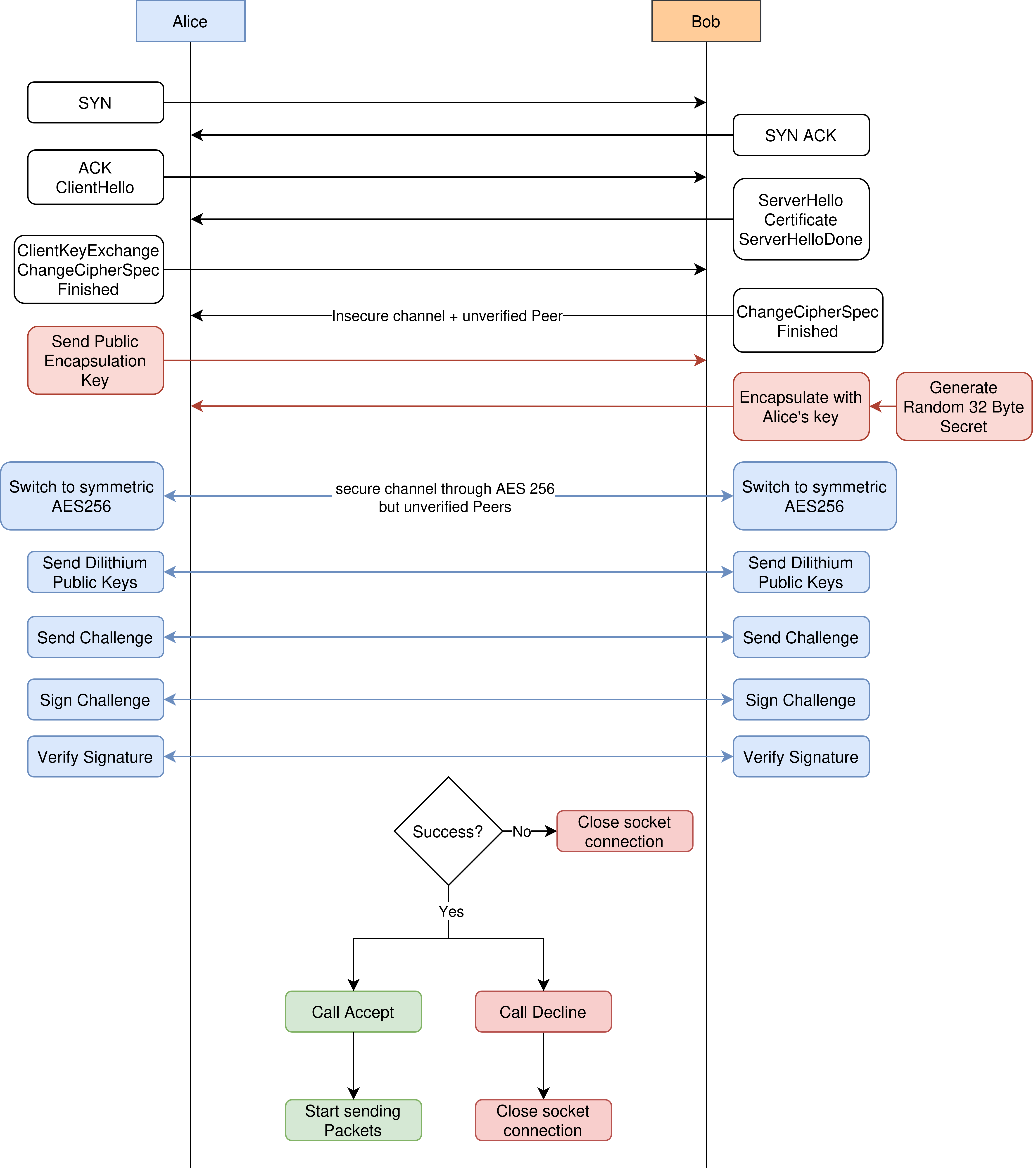}}
\caption{SAPEC modified TLS handshake. Layers correspond to OSI model levels 4, 6, and 7. 
The custom protocol layer above TLS performs mutual ML-KEM post-quantum key establishment, ephemeral key 
exchange, identity authentification, nonce synchronization, and call accept/decline signaling 
before any audio is transmitted.}
\label{fig:handshake}
\end{figure}

The handshake protocol, depicted in Fig.~\ref{fig:handshake}, constitutes the primary technical 
contribution of this work. It proceeds in three distinct layers above the TCP transport. First, a standard TLS session is established between the two endpoints, providing an authenticated and 
encrypted channel over which the subsequent custom protocol operates\cite{tlsrfc}. This TLS session is treated as an untrusted intermediate step rather than a security boundary, since its asymmetric primitives are not quantum-resistant\cite{tlsmicrosoft}.

Following TLS establishment, the endpoints perform a post-quantum key establishment using ML-KEM-1024 \cite{fips203}. Alice transmits her public encapsulation key to Bob, who performs the encapsulation operation and returns the resulting ciphertext. Alice decapsulates the received ciphertext using her corresponding private key. This process establishes an identical 32-byte shared secret at both endpoints without transmitting the secret itself in plaintext. Both parties subsequently switch to symmetric AES-256 encryption using the established session secret. At this stage, communication is protected by a quantum-resistant symmetric encryption layer, but the identities of the communicating peers have not yet been mutually verified.

Mutual identity verification is subsequently performed using the CRYSTALS-Dilithium digital signature scheme \cite{dilithium}. After the AES-256 channel has been established, both endpoints exchange their Dilithium public verification keys. Each party then sends a challenge to its counterpart. The receiving endpoint signs the challenge using its corresponding Dilithium private signing key and returns the resulting signature. The originating endpoint verifies the received signature against the previously exchanged public verification key. This challenge-signature-verification sequence is performed reciprocally by both parties. Successful verification demonstrates possession of the private signing key corresponding to the presented public identity key, thereby establishing mutual cryptographic identity without relying on the previous ML-KEM-based encryption challenge-response mechanism.

If signature verification fails, the connection is immediately terminated by closing the socket. If both parties successfully verify each other's signatures, the protocol proceeds to the call-control stage, in which the endpoints exchange call acceptance or rejection signals. An accepted call initiates audio packet transmission, while a declined call results in the socket connection being closed.

\subsection{Metadata Minimization}
\label{subsec:metadata}

Beyond the encryption of call content, SAPEC applies several complementary mechanisms to reduce 
the metadata observable by a network adversary. The most significant of these is the mandatory use of Tor hidden services for all communication. Because both endpoints are addressed by onion 
addresses rather than IP addresses, and because all traffic travels exclusively within the Tor 
network, neither party's true IP address is revealed to the other, to any relay, or to a passive observer on the public Internet\cite{toroverview}. The standard Tor circuit of three relays ensures that no single node can correlate both origin and destination.

At the packet level, audio is transmitted in fixed 20-millisecond intervals regardless of voice 
activity. During periods of silence, packets are padded with random bytes to their standard 
transmission size rather than suppressed or reduced, preventing a traffic analyst from knowing which party is speaking, when speech begins or ends, or the rhythm of the conversation from packet timing or size variations alone.

Replay attacks are mitigated by a nonce counter synchronized at call setup. Each received packet carries an incrementing sequence number; packets with a nonce outside a rolling acceptance window of approximately $\pm$100\,ms are rejected. This window accommodates the timing jitter inherent to Tor circuits while preventing an adversary from replaying captured packets at an arbitrary future time to disrupt communication\cite{replayattack}.

Finally, persistent identity key material is managed through the Crypto Ignition Key (CIK), a 
removable USB device that carries the user's long-term ML-KEM key pair, Dilithium identity authentification material and the Tor hidden service private key. CipherLink monitors for CIK insertion 
and removal via a dedicated thread. The hidden service is started only when a valid CIK is 
present; if the CIK is removed during an active call, all services and the ongoing session are 
terminated immediately. This design ensures that identity keys are never written to the device's internal storage and are not easily recoverable from a seized device in the absence of the CIK, mitigating the physical access vector described in the threat model.

\section{Implementation}
\label{sec:implementation}

SAPEC is implemented across three hardware platforms that share a common software backend. All 
platforms run on 64-bit ARM\cite{arm} Linux and communicate exclusively through the CipherLink backend described below. Development proceeded iteratively: the backend was validated on a conventional desktop before being ported to each embedded target, allowing protocol correctness to be established independently of hardware constraints.

\subsection{CipherLink}

CipherLink is a Python-based backend that implements the full SAPEC protocol stack: Tor hidden 
service management, the modified TLS handshake, ML-KEM-1024 key operations via the 
\texttt{pqcrypto} library\cite{pqcrypto}, AES-256 audio encryption, nonce counter maintenance, and CIK lifecycle management. Audio is captured and played back at configurable bitrates between 
10\,kbps and 64\,kbps using the Opus codec\cite{opus}; at 32\,kbps, call quality is subjectively 
equivalent to conventional VoIP under normal Tor circuit conditions.

A notable constraint of the Tor hidden service model influenced the socket architecture: because 
Tor does not expose the caller's return address to the called party, bidirectional audio cannot 
be established through two independent connections. Instead, both the inbound signaling channel 
and the return audio stream are multiplexed over the single circuit initiated by the caller, 
with the called party's audio socket bound to the same connection\cite{toroverview}. A local web server exposes a 
browser-based graphical interface over localhost, communicating with the backend through a 
WebSocket; this architecture allows the same backend to serve both a full desktop UI and the 
embedded interfaces described below without modification to the core protocol code.

\subsection{SAPEC Desk Phone}

The SAPEC Desk Phone embeds a Raspberry Pi 5 \cite{raspi5} inside the housing of a Tel\,01\,LX, a 
single-line desk telephone manufactured for the German Bundespost in the 1980s \cite{telefon}. The handset is 
connected to the Raspberry Pi via a USB audio interface, providing microphone input and speaker 
output through the original handset capsules. The telephone's piezoelectric buzzer is driven 
directly from a PWM-capable GPIO pin to produce ring, dial, and error tones.

The original keypad matrix uses a four-column, $n$-row arrangement with no integrated scanning 
logic. CipherLink therefore scans the matrix in a dedicated thread by sequentially asserting 
each column line and sampling all row lines, detecting key closures at a rate sufficient to 
capture input reliably during an active call. Special keys are mapped to call-control functions: 
\texttt{R} initiates a fresh Tor circuit without terminating the application, and \texttt{X} 
hangs up without requiring the handset to be returned to the cradle. A separate thread monitors 
the hook-switch state and terminates any active session when the handset is replaced. Contacts 
are stored in a local \texttt{phonebook.json} file mapping user-defined labels to onion 
addresses for testing  \cite{json}.

A USB Type-A port\cite{usb} was installed in the telephone housing and wired to an internal USB header on 
the Raspberry Pi to accommodate the CIK. A dedicated monitoring thread polls for USB device 
insertion and removal; when a device with the volume label \texttt{CIK} is detected, the 
required key files are verified before the Tor hidden service is started and inbound calls are 
accepted. Removal of the CIK at any point immediately terminates all active services and ongoing calls.

\subsection{SAPEC Pager}

The SAPEC Pager is built around a Raspberry Pi Zero\,2W running Raspbian Lite (64-bit, no 
desktop environment)\cite{raspbian}. A small LCD panel is connected over SPI \cite{spi} and a compact physical keyboard 
over I\textsuperscript{2}C \cite{i2c}. Because no framebuffer driver compatible with the 64-bit kernel was 
available for the display module, the entire user interface was implemented from scratch as a 
Python rendering layer that writes pixel data directly to the display controller's memory over SPI.

The UI provides a login screen, a scrollable contact list with shortcut-dial entries, a WLAN \cite{wlan} selection and credential interface driven through \texttt{nmcli} \cite{nmcli}, and an incoming call overlay. The system console is disabled and a \texttt{systemd} \cite{systemd} service starts the CipherLink interface automatically on boot, so the device presents no interactive shell to an unauthenticated user.

\section{Evaluation}
\label{sec:evaluation}

Measurements were conducted to characterize SAPEC's performance along four dimensions: 
connection setup time, one-way audio delivery delay, round-trip time (RTT), and audio quality. All tests were executed from a cable broadband connection over Wi-Fi on 
desktop-class hardware to simulate a probable end-user setup.

\subsection{Measurement Methodology}

A two-node testbed was configured using a purpose-built profiling harness consisting of an 
\textit{Active Peer} and a \textit{Relay Beacon}. The Active Peer runs a modified CipherLink 
backend that embeds a high-resolution timestamp into every outgoing audio packet. The Relay 
Beacon receives each packet and either returns it immediately, allowing the Active Peer to 
compute the round-trip time, or records the arrival timestamp and transmits it back to allow 
one-way delay calculation. Both nodes rebuild their Tor circuits independently after each 
measurement run, with a 10-second inter-run pause to allow circuit teardown and re-establishment and to avoid unnecessary excessive strain on the tor-network. 
Setup time is measured from the moment the Active Peer initiates a connection attempt to the 
moment the first audio packet is successfully received by the Beacon. A total of 45 independent 
test runs were executed (C1–C45).

\subsection{Connection Setup Time}

\begin{figure}[htbp]
\centerline{\includegraphics[width=\columnwidth]{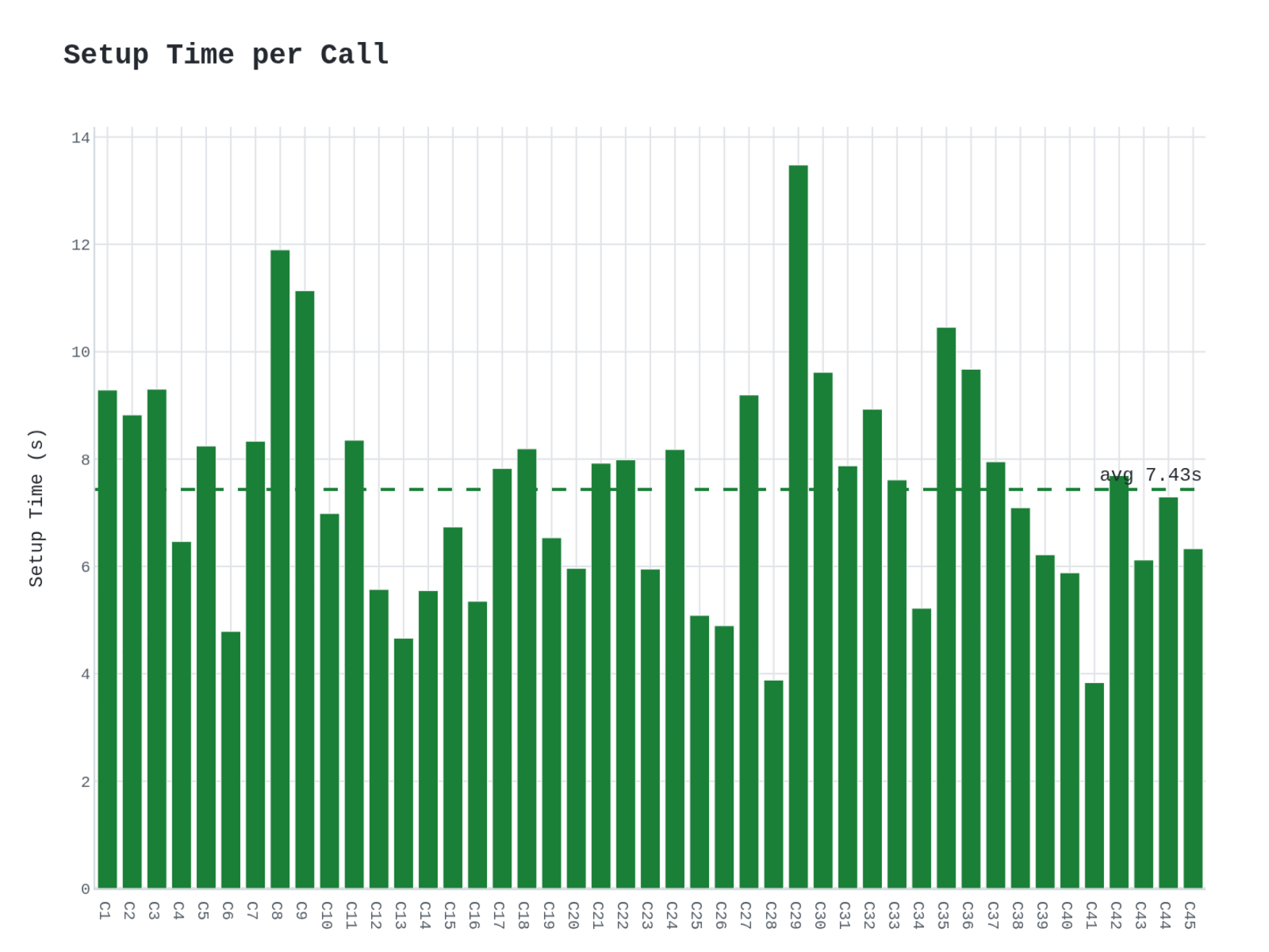}}
\caption{Connection setup time across 45 test runs. The dashed horizontal line indicates the 
mean of 7.43\,s. Three outlier runs (C9, C24, C30) correspond to weak Tor circuits nodes that had a high initial connection delay.}
\label{fig:setup}
\end{figure}

Fig.~\ref{fig:setup} shows the connection setup time across all 45 runs. Setup time subsumes Tor circuit construction, TLS negotiation, ML-KEM identity exchange, ephemeral key derivation and nonce counter synchronization. The mean setup time was 
7.43\,s, with a median of 7.61\,s and a range of 3.84\,s to 13.48\,s. The majority of runs fell between 5\,s and 10\,s, consistent with typical Tor hidden-service rendezvous latency. The three highest-latency runs correspond to circuit configurations in which Tor selected relay nodes that may have had slower relay times and/or which geographical distance was significantly larger.

\subsection{Audio Transmission Latency}

\begin{figure}[htbp]
\centerline{\includegraphics[width=\columnwidth]{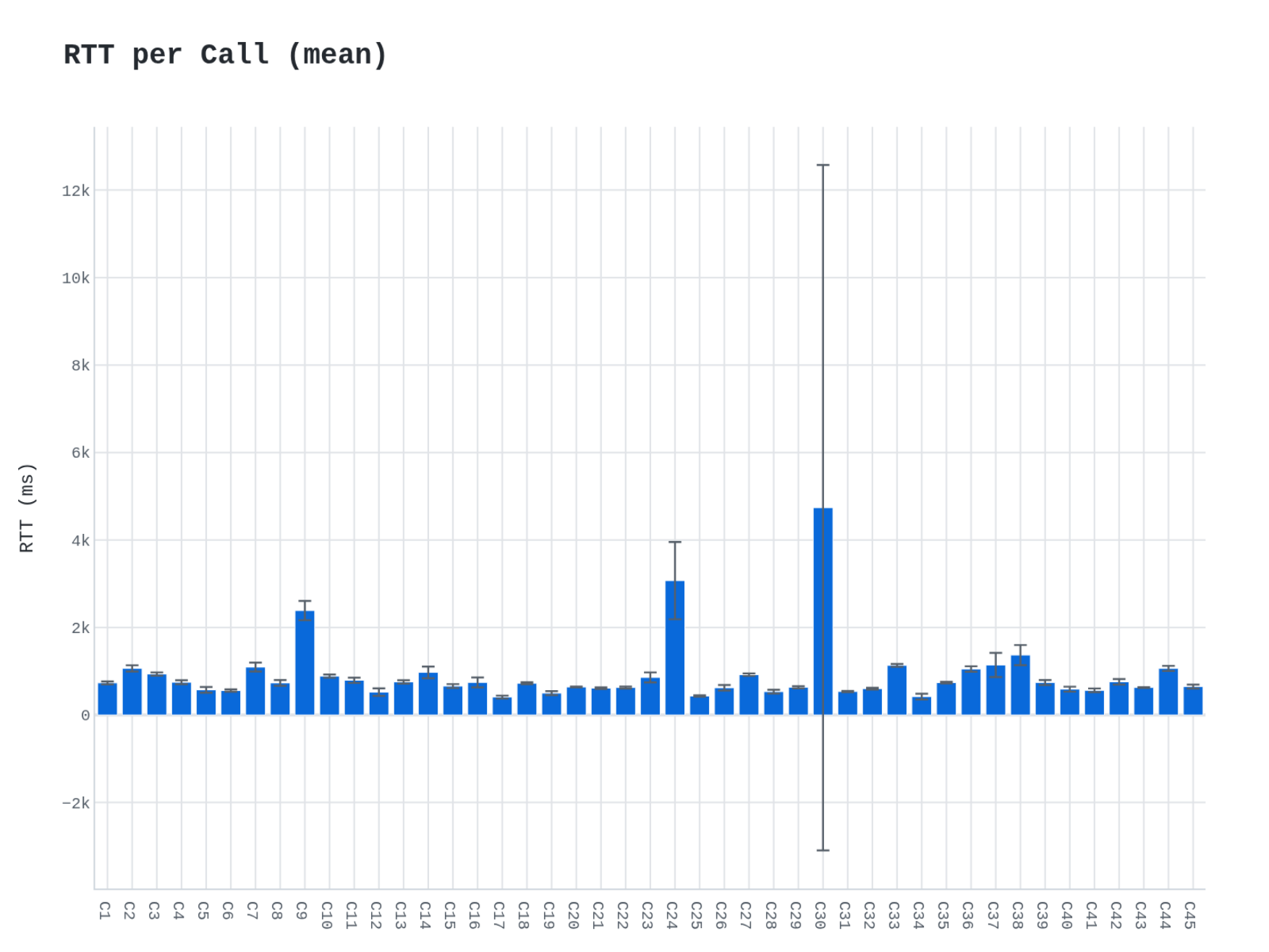}}
\caption{Round-trip time (RTT) across 45 test runs. 
Bars show per-call means; error indicators show within-call standard deviation. Calls C9, C24, and C30 are identified as outliers attributable to weak automatic Tor circuit node selection.}
\label{fig:latencyrtt}
\end{figure}

\begin{figure}[htbp]
\centerline{\includegraphics[width=\columnwidth]{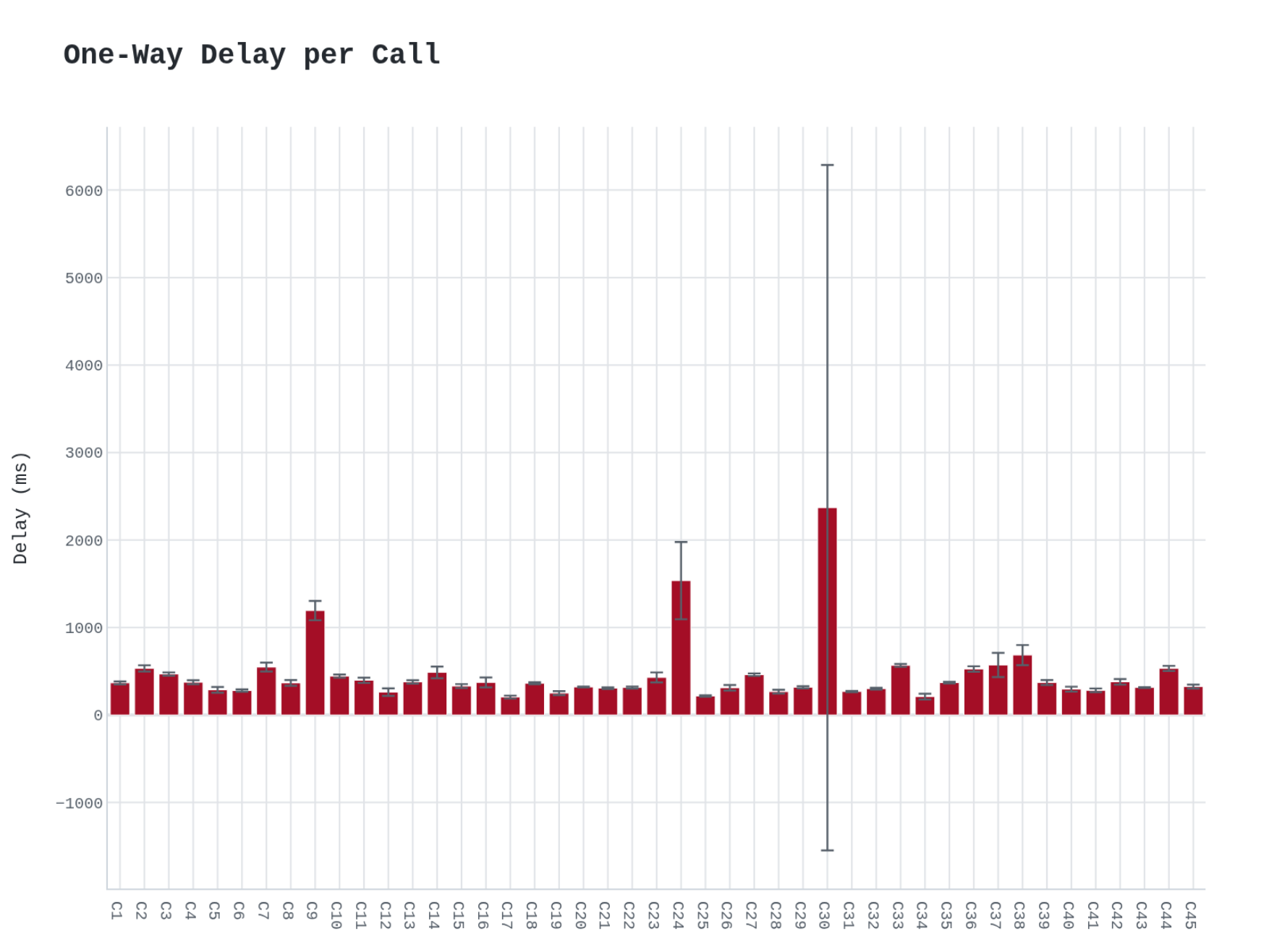}}
\caption{Distribution of per-call mean one-way delay across all 45 runs, 
excluding the three outlier calls (C9, C24, C30).}
\label{fig:latency_dist}
\end{figure}

Latency results are shown in Fig.~\ref{fig:latencyrtt} and Fig.~\ref{fig:latency_dist}. 
Table~\ref{tab:latency} summarizes the key statistics across all 45 runs and for the 42-run subset excluding the three identified outliers.

\begin{table}[htbp]
\caption{Latency Statistics Across 45 Test Runs}
\begin{center}
\begin{tabular}{lcccc}
\hline
\textbf{Metric} & \textbf{Mean} & \textbf{Median} & \textbf{Min} & \textbf{Max} \\
\hline
OWD – all runs & 460\,ms & 366\,ms & 204\,ms & 2369\,ms \\
OWD – excl. outliers & 356\,ms & 355\,ms & 204\,ms & 684\,ms \\
RTT – all runs & 921\,ms & 732\,ms & 408\,ms & 4738\,ms \\
RTT – excl. outliers & 713\,ms & 710\,ms & 408\,ms & 1368\,ms \\
\hline
\end{tabular}
\label{tab:latency}
\end{center}
\end{table}

Across the 42 non-outlier runs, the median one-way delay was 355\,ms and the median RTT was 710\,ms, both of which are attributable primarily to Tor circuit propagation rather than protocol processing overhead. The minimum observed OWD of 204\,ms represents a lower bound imposed by the Tor network architecture; no further reduction is achievable without departing from onion routing per our assessments. Within individual calls, RTT remained stable after the initial circuit-stabilization phase, as visible in the per-call time series. Calls C9, C24, and C30 exhibited anomalously high latency (OWD means of 1193\,ms, 1535\,ms, and 2369\,ms respectively) consistent with Tor circuit paths spanning over geographically larger distances and/or nodes with slower connections; these runs are retained in the dataset but identified separately.

\subsection{Audio Quality}

Audio is encoded using the Opus codec \cite{opus} at a fixed bitrate. For this test, the bitrate was set on 64 kbit/s. Given that SAPEC targets secure telephony rather than high-fidelity audio, 32\,kbps is the 
recommended operating point, as it remains well below the bandwidth available on modern broadband connection and does not increase Tor circuit stress, per the authors independent estimates.

\section{Discussion}
\label{sec:discussion}

\subsection{Interpretation of Results}

The evaluation demonstrates that real-time voice communication over Tor hidden services is practically achievable. Excluding the three outlier runs attributable to unfavorable automatic Tor circuit selection, the median one-way delay of 355\,ms and median RTT of 710\,ms are consistent with the latency profiles reported for other real-time applications routed over Tor \cite{torlatency}. While these figures partly exceed the 300\,ms OWD threshold recommended by ITU-T G.114\cite{ITU} for interactive telephony, users in informal testing found the resulting call quality acceptable for secure communications where privacy is prioritized over low latency. The stable within-call RTT observed across the majority of runs indicates that, once a Tor circuit is established, it does not introduce significant additional jitter during the call itself. The three outlier runs (C9, C24, C30) illustrate the non-deterministic nature of Tor circuit construction: when relay selection spans geographically distant nodes or nodes with limited bandwidth, latency increases substantially. The 10-second inter-run circuit rebuild in the test harness did not deterministically prevent such selections, suggesting that selective circuit management could improve consistency in a production deployment. This partly undermines the security design of the tor network \cite{toroverview} and is therefore not implemented in our current prototype.

\section{Conclusion}
\label{sec:conclusion}

This paper presented SAPEC, a peer-to-peer voice communication system that combines 
post-quantum key encapsulation via ML-KEM-1024, anonymous overlay routing via Tor hidden 
services, identity verification with Dilithium and a custom protocol layer above TLS to provide simultaneously anonymous, 
decentralized, and quantum-resistant telephony. To the best of our knowledge, SAPEC is the first system to realize all three properties in a working voice communication prototype on consumer embedded hardware. Empirical evaluation across 45 test runs demonstrated a median connection setup time of 7.61\,s and a median one-way audio delay of 355\,ms on same-region Tor circuits where the latency is partly attributable to the anonymization layer rather than to the cryptographic protocol. The prototype has been realized in three form factors: a desktop software client, a retrofitted 1980s desk telephone, and a portable pager, demonstrating the portability of the CipherLink backend across constrained embedded targets.

\section{Acknowledgments \& Disclaimer}
We want to express special thanks to \mbox{\textbf{Hans Eberhardt Pulter}} who was the supervisor of this project and helped with logistics as well as sourcing of materials and supported us with his expertise in working with electronics and hardware.
\newline

This Paper was written with language related assistance from various AI models due to english not being the first language of the authors. Everything written is fact checked and reviewed by the authors to the best of their knowledge.

\end{document}